\documentclass[a4paper,11pt]{article}

\makeatletter
\gdef\@fpheader{   }
\gdef\@journal{jhep}

\RequirePackage{amsmath}
\RequirePackage{amssymb}
\RequirePackage{epsfig}
\RequirePackage{CJK}
\RequirePackage{graphicx}
\RequirePackage[numbers,sort&compress]{natbib}
\RequirePackage{color}
\RequirePackage[%dvipdfmx
,colorlinks=true
,urlcolor=blue
,anchorcolor=blue
,citecolor=blue
,filecolor=blue
,linkcolor=blue
,menucolor=blue
,pagecolor=blue
,linktocpage=true
,pdfproducer=medialab
,pdfa=true
,CJKbookmarks
]{hyperref}

\newif\ifnotoc\notocfalse
\newif\ifemailadd\emailaddfalse
\newif\iftoccontinuous\toccontinuousfalse

\def\@subheader{\@empty}
\def\@keywords{\@empty}
\def\@abstract{\@empty}
\def\@xtum{\@empty}
\def\@dedicated{\@empty}
\def\@arxivnumber{\@empty}
\def\@collaboration{\@empty}
\def\@collaborationImg{\@empty}
\def\@proceeding{\@empty}
\def\@preprint{\@empty}

\newcommand{\subheader}[1]{\gdef\@subheader{#1}}
\newcommand{\keywords}[1]{\if!\@keywords!\gdef\@keywords{#1}\else%
\PackageWarningNoLine{\jname}{Keywords already defined.\MessageBreak Ignoring last definition.}\fi}
\renewcommand{\abstract}[1]{\gdef\@abstract{#1}}
\newcommand{\dedicated}[1]{\gdef\@dedicated{#1}}
\newcommand{\arxivnumber}[1]{\gdef\@arxivnumber{#1}}
\newcommand{\proceeding}[1]{\gdef\@proceeding{#1}}
\newcommand{\xtumfont}[1]{\textsc{#1}}
\newcommand{\correctionref}[3]{\gdef\@xtum{\xtumfont{#1} \href{#2}{#3}}}
\newcommand\jname{JHEP}
\newcommand\acknowledgments{\section*{Acknowledgments}}

\newcommand\preprint[1]{\gdef\@preprint{\hfill #1}}

\newcommand\note[2][]{%
\if!#1!%
\stepcounter{footnote}\footnotetext{#2}%
\else%
{\renewcommand\thefootnote{#1}%
\footnotetext{#2}}%
\fi}

\newtoks\auth@toks
\renewcommand{\author}[2][]{%
  \if!#1!%
    \auth@toks=\expandafter{\the\auth@toks#2\ }%
  \else
    \auth@toks=\expandafter{\the\auth@toks#2$^{#1}$\ }%
  \fi
}

\newtoks\affil@toks\newif\ifaffil\affilfalse
\newcommand{\affiliation}[2][]{%
\affiltrue
  \if!#1!%
    \affil@toks=\expandafter{\the\affil@toks{\item[]#2}}%
  \else
    \affil@toks=\expandafter{\the\affil@toks{\item[$^{#1}$]#2}}%
  \fi
}

\newtoks\email@toks\newcounter{email@counter}%
\newcommand{\emailAdd}[1]{%
\emailaddtrue%
\ifnum\theemail@counter>0\email@toks=\expandafter{\the\email@toks, \@email{#1}}%
\else\email@toks=\expandafter{\the\email@toks\@email{#1}}%
\fi\stepcounter{email@counter}}
\newcommand{\@email}[1]{\href{mailto:#1}{\tt #1}}

\newcommand*\collaboration[1]{\gdef\@collaboration{#1}}
\newcommand*\collaborationImg[2][]{\gdef\@collaborationImg{#2}}

\newcommand\afterLogoSpace{\smallskip}
\newcommand\afterSubheaderSpace{\vskip3pt plus 2pt minus 1pt}
\newcommand\afterProceedingsSpace{\vskip21pt plus0.4fil minus15pt}
\newcommand\afterTitleSpace{\vskip23pt plus0.06fil minus13pt}
\newcommand\afterRuleSpace{\vskip23pt plus0.06fil minus13pt}
\newcommand\afterCollaborationSpace{\vskip3pt plus 2pt minus 1pt}
\newcommand\afterCollaborationImgSpace{\vskip3pt plus 2pt minus 1pt}
\newcommand\afterAuthorSpace{\vskip5pt plus4pt minus4pt}
\newcommand\afterAffiliationSpace{\vskip3pt plus3pt}
\newcommand\afterEmailSpace{\vskip16pt plus9pt minus10pt\filbreak}
\newcommand\afterXtumSpace{\par\bigskip}
\newcommand\afterAbstractSpace{\vskip16pt plus9pt minus13pt}
\newcommand\afterKeywordsSpace{\vskip16pt plus9pt minus13pt}
\newcommand\afterArxivSpace{\vskip3pt plus0.01fil minus10pt}
\newcommand\afterDedicatedSpace{\vskip0pt plus0.01fil}
\newcommand\afterTocSpace{\bigskip\medskip}
\newcommand\afterTocRuleSpace{\bigskip\bigskip}
\newlength{\affiliationsSep}
\newcommand\beforetochook{\pagestyle{myplain}\pagenumbering{roman}}

\DeclareFixedFont\trfont{OT1}{phv}{b}{sc}{11}

\renewcommand\maketitle{
\pagestyle{empty}
\thispagestyle{titlepage}
\setcounter{page}{0}
\noindent{\small\scshape\@fpheader}\@preprint\par
\afterLogoSpace
\if!\@subheader!\else\noindent{\trfont{\@subheader}}\fi
\afterSubheaderSpace
\if!\@proceeding!\else\noindent{\sc\@proceeding}\fi
\afterProceedingsSpace
{\LARGE\flushleft\sffamily\bfseries\@title\par}
\afterTitleSpace
\hrule height 1.5\p@%
\afterRuleSpace
\if!\@collaboration!\else
{\Large\bfseries\sffamily\raggedright\@collaboration}\par
\afterCollaborationSpace
\fi
\if!\@collaborationImg!\else
{\normalsize\bfseries\sffamily\raggedright\@collaborationImg}\par
\afterCollaborationImgSpace
\fi
{\bfseries\raggedright\sffamily\the\auth@toks\par}
\afterAuthorSpace
\ifaffil\begin{list}{}{%
\setlength{\leftmargin}{0.28cm}%
\setlength{\labelsep}{0pt}%
\setlength{\itemsep}{\affiliationsSep}%
\setlength{\topsep}{-\parskip}}
\itshape\small%
\the\affil@toks
\end{list}\fi
\afterAffiliationSpace
\ifemailadd %% if emailadd is true
\noindent\hspace{0.28cm}\begin{minipage}[l]{.9\textwidth}
\begin{flushleft}
\textit{E-mail:} \the\email@toks
\end{flushleft}
\end{minipage}
\else %% if emailaddfalse do nothing
\PackageWarningNoLine{\jname}{E-mails are missing.\MessageBreak Plese use \protect\emailAdd\space macro to provide e-mails.}
\fi
\afterEmailSpace
\if!\@xtum!\else\noindent{\@xtum}\afterXtumSpace\fi
\if!\@abstract!\else\noindent{\renewcommand\baselinestretch{.9}\textsc{Abstract:}}\ \@abstract\afterAbstractSpace\fi
\if!\@keywords!\else\noindent{\textsc{Keywords:}} \@keywords\afterKeywordsSpace\fi
\if!\@arxivnumber!\else\noindent{\textsc{ArXiv ePrint:}} \href{http://arxiv.org/abs/\@arxivnumber}{\@arxivnumber}\afterArxivSpace\fi
\if!\@dedicated!\else\vbox{\small\it\raggedleft\@dedicated}\afterDedicatedSpace\fi
\ifnotoc\else
\iftoccontinuous\else\newpage\fi
\beforetochook\hrule
\tableofcontents
\afterTocSpace
\hrule
\afterTocRuleSpace
\fi
\setcounter{footnote}{0}
\pagestyle{myplain}\pagenumbering{arabic}
} % close the \renewcommand\maketitle{

\renewcommand{\baselinestretch}{1.1}\normalsize
\divide\@tempdima\baselineskip
\@tempcnta=\@tempdima
\skip\@mpfootins = \skip\footins
\renewcommand{\@dotsep}{10000}

\newcommand\ps@myplain{
\pagenumbering{arabic}
\renewcommand\@oddfoot{\hfill-- \thepage\ --\hfill}
\renewcommand\@oddhead{}}
\let\ps@plain=\ps@myplain

\newcommand\ps@titlepage{\renewcommand\@oddfoot{}\renewcommand\@oddhead{}}

\numberwithin{equation}{section}

\renewcommand\section{\@startsection{section}{1}{\z@}%
                                   {-3.5ex \@plus -1.3ex \@minus -.7ex}%
                                   {2.3ex \@plus.4ex \@minus .4ex}%
                                   {\normalfont\large\bfseries}}
\renewcommand\subsection{\@startsection{subsection}{2}{\z@}%
                                   {-2.3ex\@plus -1ex \@minus -.5ex}%
                                   {1.2ex \@plus .3ex \@minus .3ex}%
                                   {\normalfont\normalsize\bfseries}}
\renewcommand\subsubsection{\@startsection{subsubsection}{3}{\z@}%
                                   {-2.3ex\@plus -1ex \@minus -.5ex}%
                                   {1ex \@plus .2ex \@minus .2ex}%
                                   {\normalfont\normalsize\bfseries}}
\renewcommand\paragraph{\@startsection{paragraph}{4}{\z@}%
                                   {1.75ex \@plus1ex \@minus.2ex}%
                                   {-1em}%
                                   {\normalfont\normalsize\bfseries}}
\renewcommand\subparagraph{\@startsection{subparagraph}{5}{\parindent}%
                                   {1.75ex \@plus1ex \@minus .2ex}%
                                   {-1em}%
                                   {\normalfont\normalsize\bfseries}}

\def\fnum@figure{\textbf{\figurename\nobreakspace\thefigure}}
\def\fnum@table{\textbf{\tablename\nobreakspace\thetable}}

\long\def\@makecaption#1#2{%
  \vskip\abovecaptionskip
  \sbox\@tempboxa{\small #1. #2}%
  \ifdim \wd\@tempboxa >\hsize
    \small #1. #2\par
  \else
    \global \@minipagefalse
    \hb@xt@\hsize{\hfil\box\@tempboxa\hfil}%
  \fi
  \vskip\belowcaptionskip}

\renewenvironment{thebibliography}[1]{%
\begin{oldthebibliography}{#1}%
\small%
\raggedright%
\setlength{\itemsep}{5pt plus 0.2ex minus 0.05ex}%
}%
{%
\end{oldthebibliography}%
}

\makeatother

\usepackage[T1]{fontenc} % if needed
\usepackage{romannum} 
\usepackage{CJK}

\title{{\boldmath Toy models of black hole, white hole and wormhole: thermal effects and information loss problem}}
\author[a,1]{Yu-Zhu Chen,\note{chenyuzhu@nankai.edu.cn}}
\author[b]{Shi-Lin Li, }
\author[b]{Yu-Jie Chen, }
\author[b]{Fu-Lin Zhang, }
\author[b,2]{Wu-Sheng Dai \note{daiwusheng@tju.edu.cn}}

\affiliation[a]{Theoretical Physics Division, Chern Institute of Mathematics, Nankai University, Tianjin 300071, P.R. China}
\affiliation[b]{Department of Physics, Tianjin University, Tianjin 300072, P.R. China}

\abstract{In this paper, by setting proper boundaries in the Minkowski spacetime, we
construct three toy model spacetimes, a toy black hole, a toy white hole, and
a toy wormhole. Based on these model spacetimes, we discuss the
Hawking radiation and the information loss problem. By counting the
number of the field modes inside and outside the horizon, we show the thermal
radiation of the toy black hole. We show that the white hole have a thermal
absorption. We show that in the whole toy wormhole spacetime, there is no
information lost. In addition, we show the black hole radiation and the white
hole absorption are independent of the choices of boundary conditions at the
singularity. We also show the physical effects caused by two particular
boundary conditions.
}


\begin{document} %正文开始
\begin{CJK*}{GBK}{song}
\maketitle %生成题目

\flushbottom
%(正文开始) ――――――――――――――――――――――――――――――――――――――――――――――――――
\section{Introduction}

After Hawking revealed the Hawking radiation by considering the scalar field
in a collapsing black hole \cite{hawking1975particle}, the Hawking radiation
are calculated for various black hole systems and various kinds of particles,
such as the Hawking radiation of a spherical collapsing shell and of the
Kerr-Newmann black hole
\cite{boulware1976hawking,davies1976origin,zhang2005new}\textbf{, }the Hawking
radiation in the Ho\v{r}ava-Lifshitz gravity\textbf{ }\cite{majhi2010hawking}%
\textbf{, }the Hawking radiation in the Einstein-dilaton-Gauss-Bonnet black
hole\textbf{ }\cite{konoplya2019quasinormal}, the Hawking radiation of the
Dirac particle in the Kerr black hole via tunneling \cite{li2008hawking}, the
Hawking radiation of spin-1 particles in a rotating hairy black hole and
spin-2 particles in a static spherical black hole
\cite{sakalli2015hawking,sakalli2016black}, and the Hawking radiation of
type-II Weyl fermions \cite{volovik2016black}. The Hawking radiation is also
verified by different methods. The density matrix method of the Hawking
radiation is used in ref.
\cite{parker1975probability,wald1975particle,hawking1976black}. A
path-integral calculation of the Hawking radiation is developed by Hawking and
Hartle \cite{hartle1976path}. Based on the discontinuous jump of the wave
function on the horizon, Ruffini and Damour provide a method of the Hawking
radiation for the Schwarzschild black hole \cite{damour1976black}, which is
also called the tunneling mechanism \cite{banerjee2009hawking}. The Hawking
radiation observation is considered in an analogue black hole system
\cite{steinhauer2014observation,steinhauer2016observation,de2019observation},
e.g., in Bose-Einstein condensates \cite{macher2009black}. The Hawking
radiation may cause the information loss problem. More details about the
information loss problem can be found in refs.
\cite{hawking2015information,polchinski2015new}.

Nevertheless, the universal property of the quantum field in a black hole
spacetime deserves more discussions. In Hawking's scheme the spacetime
background is complicated, so in stead of an exact result we usually consider
the asymptotic behavior of a quantum field at the horizon and at the null
infinity. In our previous work \cite{li2019scattering}, we provide an exact
solution of both scattering states and bound states of scalar fields in the
Schwarzschild spacetime. We also provide a new method for the scalar
scattering problem in the Schwarzschild spacetime \cite{li2018scalar}. In the
Schwarzschild case, however, we have to deal with the complicated special function.

In this paper, we construct toy models of the black hole and the white hole.
The black (white) hole is a universal geometry structure of spacetime, which
is irrelevant to the local details of the metric of spacetime. By setting
proper boundaries in the Minkowski spacetime, we construct a black (white)
hole. The advantage of toy models is that the spacetime is simple enough to
calculate the quantum field exactly. In the model, the geometry is simple, but
the universal property of the spacetime is nontrivial. This allows us to
concentrate only on the universal property of the spacetime. We show that the
Hawking radiation is determined by the universal property of the spacetime
rather than the local property of geometry. Meanwhile, we show there is a
thermal absorption in the white hole model. Furthermore, we consider a
spacetime in which there exist a wormhole. In addition, we compare the effects
caused by two different boundary conditions at the boundary.

In section 2, we build a toy black hole model and show the Hawking radiation.
In section 3, we build a toy white hole model and show the thermal absorption.
In section 4, we build a wormhole model and discuss the information loss
problem. In section 5, taking the toy black hole as an example, we compare the
in-going and out-going modes inside the black hole for different boundary
conditions at the singularity. The conclusion is presented in section 6.

\section{Constructing black hole in Minkowski spacetime}

\subsection{Black hole}

In this section, we build a toy black hole in the Minkowski spacetime.

The black hole is defined as a region \cite{wald2010general}
\begin{equation}
B=M-J^{-}\left(  \mathcal{J}^{+}\right)  , \label{Def.BH}%
\end{equation}
where $M$ is the whole spacetime, $\mathcal{J}^{+}$ is the future null
infinity, and $J^{-}\left(  \mathcal{J}^{+}\right)  $ is the causal past of
$\mathcal{J}^{+}$.

According to the definition (\ref{Def.BH}), we construct a toy black hole by
choosing a proper boundary in the Minkowski spacetime.

For simplicity, we consider the 1+1-dimensional case. The black hole is
constructed by setting a boundary in a 1+1-dimensional Minkowski spacetime
\begin{equation}
ds^{2}=-dt^{2}+dx^{2}. \label{Metric}%
\end{equation}
The boundary locates at
\begin{equation}
t=\sqrt{x^{2}+\eta^{2}},\text{ }t>0, \label{BH.Bound}%
\end{equation}
as illustrated in fig. 1. By the definition (\ref{Def.BH}), the region II in
fig. 1 is a black hole. The boundary (\ref{BH.Bound}) is the spacetime singularity.

Next, we compare fig. 1 with the Schwarzschild spacetime, fig. 2.

In the Schwarzschild spacetime, fig. 2, the region II, $r<2M$, is the black
hole area. The region I, $r>2M$, is the area outside the black hole. The
boundary between the regions I and II is the event horizon given by $r=2M$.
$u$ and $v$ are the Kruskal coordinates \cite{ohanian2013gravitation}.

\begin{figure}[h]
\centering
\includegraphics[width=9.0cm,height=6.05cm]{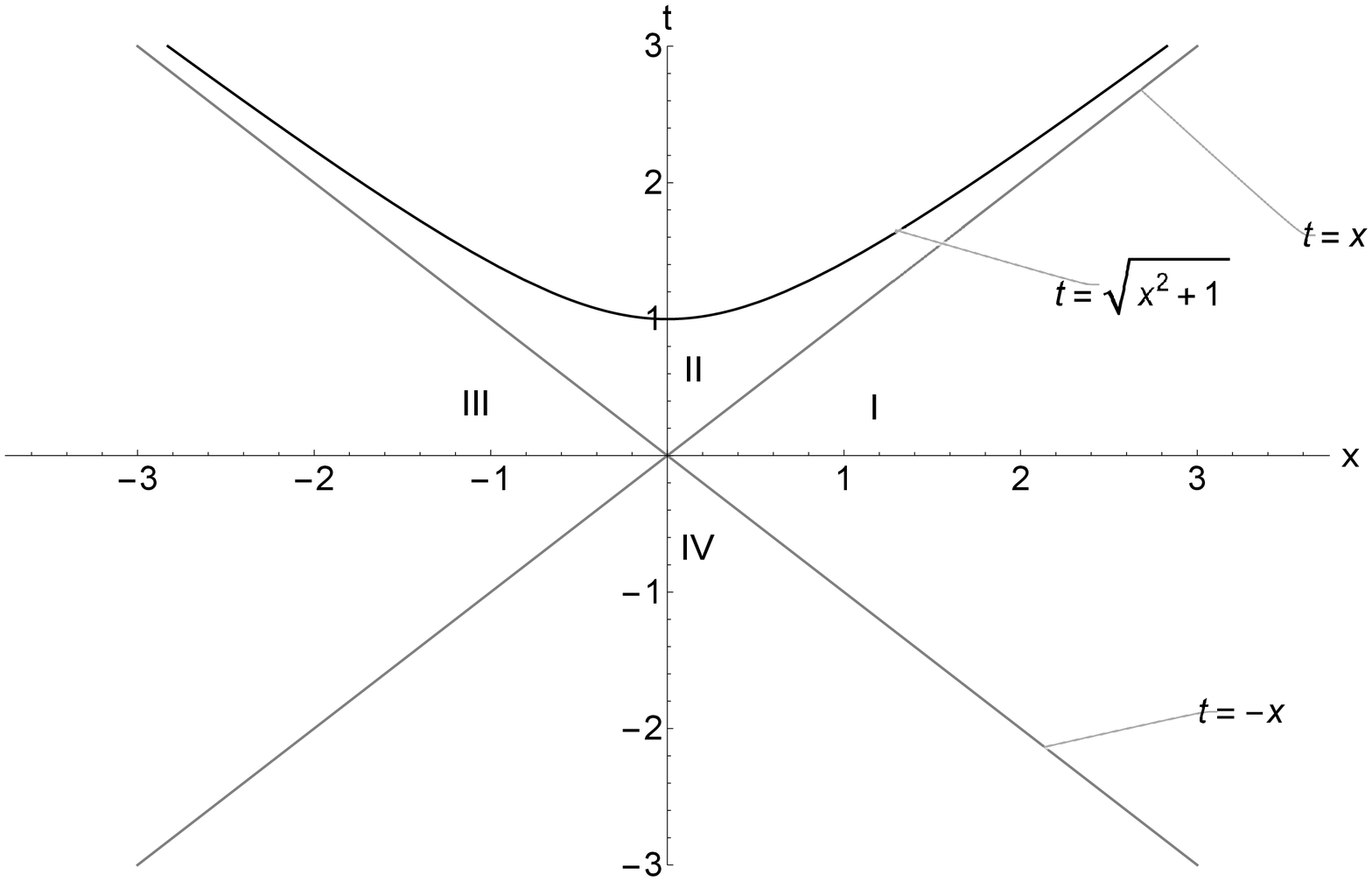}
\caption{A toy black hole}
\end{figure}

\begin{figure}[h]
\centering
\includegraphics[width=9.0cm,height=6.0cm]{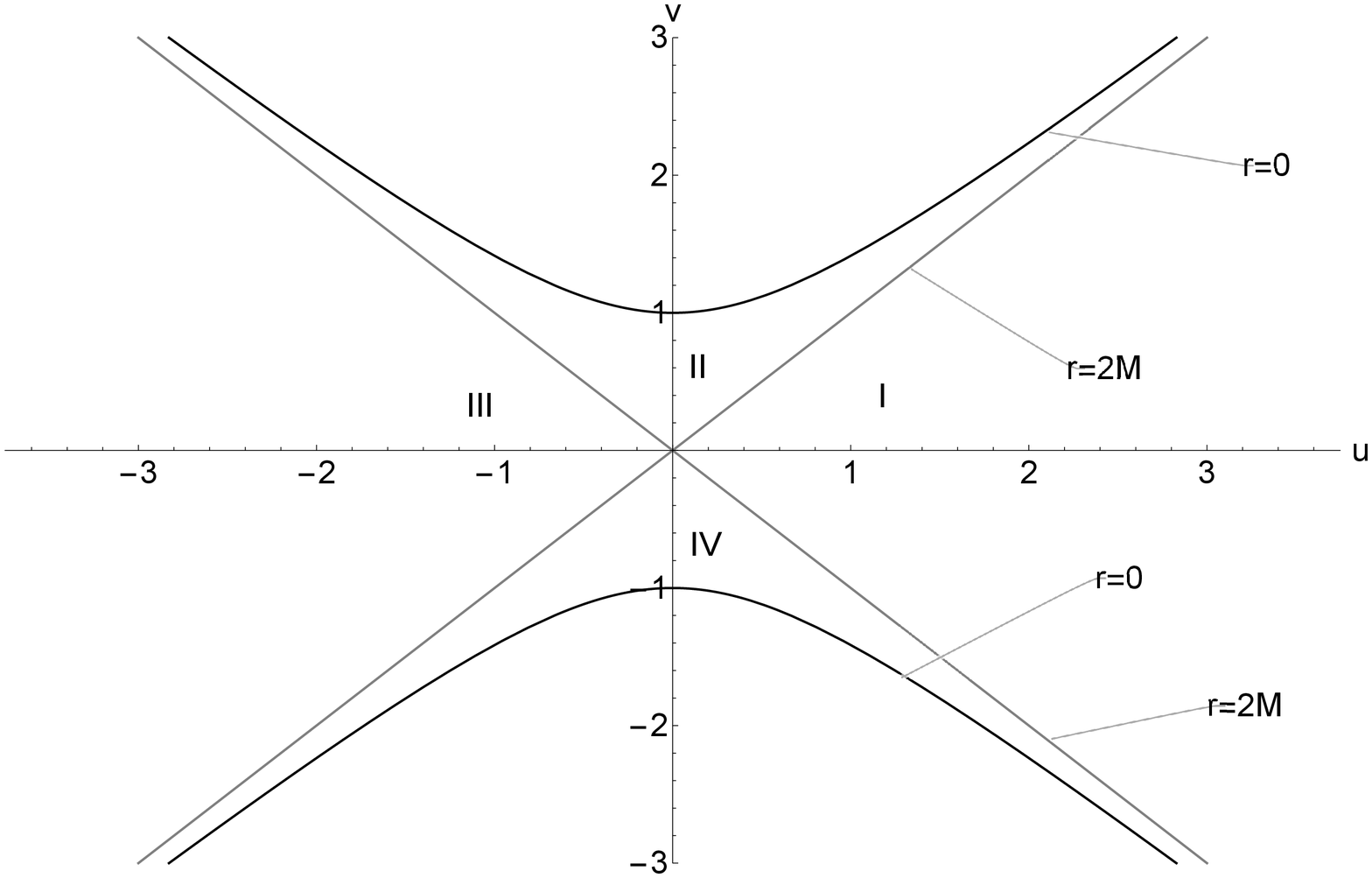}
\caption{The maximal Schwarzschild geometry}
\end{figure}

\subsection{Thermal radiation}

In this section, we consider the Hawking radiation of the black hole
constructed above.

In order to consider the Hawking radiation in the spacetime constructed above,
we consider a massless scalar field $\phi$
\begin{equation}
\left(  -\frac{\partial^{2}}{\partial t^{2}}+\frac{\partial^{2}}{\partial
x^{2}}\right)  \phi=0 \label{Massless.KG}%
\end{equation}
in the black hole spacetime. The general solution of eq. (\ref{Massless.KG})
is
\begin{equation}
\phi=f\left(  t-x\right)  +h\left(  t+x\right)  , \label{Solution.general}%
\end{equation}
where $f$ and $h$ are arbitrary functions representing the out-going modes and
in-going modes, respectively.

The region II in fig. 1 is a region with $t-x>0$ and $t+x>0$. In the region II
we introduce new coordinates,
\begin{align}
t  &  =r\cosh\theta,\nonumber\\
x  &  =r\sinh\theta. \label{BH.coordinate}%
\end{align}
Then the boundary (\ref{BH.Bound}) becomes%
\begin{equation}
r=\eta. \label{BH.Boundary}%
\end{equation}
We choose the boundary condition
\begin{equation}
\left.  \frac{\partial\phi}{\partial r}\right\vert _{r=\eta}=0 \label{BH.BC.2}%
\end{equation}
for the boundary (\ref{BH.Bound}) or (\ref{BH.Boundary}).

Using the coordinate (\ref{BH.coordinate}) and substituting eq.
(\ref{Solution.general}) into the boundary condition (\ref{BH.BC.2}), we
arrive at%
\begin{equation}
h^{\prime}\left(  \eta e^{\theta}\right)  =-e^{-2\theta}f^{\prime}\left(  \eta
e^{-\theta}\right)  . \label{assistant.1}%
\end{equation}
Integrating eq. (\ref{assistant.1}) over $\eta e^{\theta}$,%
\begin{align}
\int_{\theta=\theta_{0}}^{\theta}d\left(  \eta e^{\theta}\right)  h^{\prime
}\left(  \eta e^{\theta}\right)   &  =-\int_{\theta=\theta_{0}}^{\theta
}d\left(  \eta e^{\theta}\right)  e^{-2\theta}f^{\prime}\left(  \eta
e^{-\theta}\right)  ,\nonumber\\
&  =\int_{\theta=\theta_{0}}^{\theta}d\left(  \eta e^{-\theta}\right)
f^{\prime}\left(  \eta e^{-\theta}\right)  ,
\end{align}
we have
\begin{equation}
h\left(  \eta e^{\theta}\right)  =f^{\prime}\left(  \eta e^{-\theta}\right)
+C
\end{equation}
with $C$ the integration constant. By eq. (\ref{BH.coordinate}) we have%
\begin{align}
h\left(  \eta\sqrt{\frac{t-x}{t+x}}\right)   &  =f\left(  \eta\sqrt{\frac
{t+x}{t-x}}\right)  +C,\nonumber\\
h\left(  z\right)   &  =f\left(  \frac{\eta}{z^{2}}\right)  +C,
\end{align}
where $z\equiv\eta\sqrt{\frac{t-x}{t+x}}>0$. Without loss of generality, we
choose $C=0$. This implies that the boundary condition (\ref{BH.BC.2})
requires
\begin{equation}
f\left(  z\right)  =f\left(  \frac{\eta}{z^{2}}\right)  ,\text{ \ }z>0.
\label{relation.BH.BC.2}%
\end{equation}

In the region II $t-x>0$ and $t+x>0$, so $f\left(  t-x\right)  $ and $h\left(
t+x\right)  $ must satisfy the relation (\ref{relation.BH.BC.2}). This means
that each out-going mode $f\left(  t-x\right)  $ equals to an in-going mode
$h\left(  t+x\right)  =f\left(  \frac{\eta^{2}}{t+x}\right)  $. When $f\left(
t-x\right)  =0$, we have $h\left(  t+x\right)  =f\left(  \frac{\eta^{2}}%
{t+x}\right)  =0$.

Conversely, in the region I $t+x>0$ and $t-x<0$, so the in-going mode
$h\left(  t+x\right)  $ is restrained by the relation (\ref{relation.BH.BC.2})
while the out-going mode $f\left(  t-x\right)  $ is no longer restrained by
the relation (\ref{relation.BH.BC.2}). That is, in the region I, even though
$h\left(  t+x\right)  =0$, $f\left(  t-x\right)  $ may still be nonvanishing.
This means that the number of the out-going modes is more than the number of
the in-going modes. The extra part of the out-going modes can only exist in
the region I. For the extra part of the out-going modes, it is natural to
choose the coordinate
\begin{align}
\tau &  =\frac{1}{a}\operatorname{arctanh}\frac{t}{x},\nonumber\\
\xi &  =\frac{1}{a}\ln\left(  a\sqrt{x^{2}-t^{2}}\right)  .
\end{align}
It is coincident that $\left(  \tau,\xi\right)  $ are the comoving coordinate
of the uniformly accelerated observer with a proper acceleration $a$. For an
observer at the horizon, $a$ is the surface gravity at the horizon. By the
Bogolyubov transformation method for the calculation of the Unruh effect and
the Hawking radiation \cite{mukhanov2007introduction}, we find that the extra
modes in the region I satisfy the thermal distribution. The extra modes come
from the horizon and go to the future null infinite, which is the physical
picture of the Hawking radiation.

For completeness, we present the calculation. In the region I, the out-going
modes of the scalar field can be expressed as in the reference system $\left(
t,x\right)  $ \cite{mukhanov2007introduction,susskind2005introduction},
\begin{equation}
\hat{\phi}=\int_{0}^{\infty}\frac{d\omega}{\sqrt{2\pi}}\frac{1}{\sqrt{2\omega
}}\left(  \hat{a}_{\omega}e^{-i\omega\left(  t-x\right)  }+\hat{a}_{\omega
}^{\dagger}e^{i\omega\left(  t-x\right)  }\right)  , \label{BH.mode.expand.1}%
\end{equation}
where $\hat{a}_{\omega}$ and $\hat{a}_{\omega}^{\dagger}$ are the annihilation
and creation operators of the frequency $\omega$. The out-going modes of the
scalar field can be also expressed as in the reference system $\left(
\tau,\xi\right)  $,
\begin{equation}
\hat{\phi}=\int_{0}^{\infty}\frac{d\Omega}{\sqrt{2\pi}}\frac{1}{\sqrt{2\Omega
}}\left(  \hat{b}_{\Omega}e^{-i\Omega\left(  \tau-\xi\right)  }+\hat
{b}_{\Omega}^{\dagger}e^{i\Omega\left(  \tau-\xi\right)  }\right)  ,
\label{BH.mode.expand.2}%
\end{equation}
where $\hat{b}_{\Omega}$ and $\hat{b}_{\Omega}^{\dagger}$ are the annihilation
and creation operators of the frequency $\Omega$. Eqs. (\ref{BH.mode.expand.1}%
) and (\ref{BH.mode.expand.2}) both are the out-going modes, so they must be
equal to each other:
\begin{align}
&  \int_{0}^{\infty}\frac{d\omega}{\sqrt{2\pi}}\frac{1}{\sqrt{2\omega}}\left(
\hat{a}_{\omega}e^{-i\omega\left(  t-x\right)  }+\hat{a}_{\omega}^{\dagger
}e^{i\omega\left(  t-x\right)  }\right) \nonumber\\
&  =\int_{0}^{\infty}\frac{d\Omega}{\sqrt{2\pi}}\frac{1}{\sqrt{2\Omega}%
}\left(  \hat{b}_{\Omega}e^{-i\Omega\left(  \tau-\xi\right)  }+\hat{b}%
_{\Omega}^{\dagger}e^{i\Omega\left(  \tau-\xi\right)  }\right)  .
\label{BH.mode.expand}%
\end{align}
The operators $\hat{a}_{\omega}$ ($\hat{a}_{\omega}^{\dagger}$) and $\hat
{b}_{\Omega}$ ($\hat{b}_{\Omega}^{\dagger}$) are connected by the Bogolyubov
transformation
\begin{equation}
\hat{b}_{\Omega}=\int_{0}^{\infty}d\omega\left(  \alpha_{\Omega\omega}\hat
{a}_{\omega}-\beta_{\Omega\omega}\hat{a}_{\omega}^{\dagger}\right)
\end{equation}
with
\begin{equation}
\alpha_{\Omega\omega}\left(  \beta_{\Omega\omega}\right)  =\pm\frac{1}{2\pi
a}\sqrt{\frac{\Omega}{\omega}}e^{\pm\frac{\pi\Omega}{2a}}e^{\frac{i\Omega}%
{a}\ln\frac{\omega}{a}}\Gamma\left(  -\frac{i\Omega}{a}\right)  .
\label{coefficient}%
\end{equation}
By the annihilation operators $\hat{a}_{\omega}$ and $\hat{b}_{\Omega}$, we
can define two vacuum states: the Minkowski vacuum state $\left\vert
0_{M}\right\rangle $ in the reference system $\left(  t,x\right)  $:
\begin{equation}
\hat{a}_{\omega}\left\vert 0_{M}\right\rangle =0
\end{equation}
and the Rindler vacuum state $\left\vert 0_{R}\right\rangle $\ in the
reference system $\left(  \tau,\xi\right)  $:%
\begin{equation}
\hat{b}_{\Omega}\left\vert 0_{R}\right\rangle =0.
\end{equation}
We call the particle created by the operator $\hat{a}_{\omega}^{\dagger}$ a
Minkowski particle and the particle created by the operator $\hat{b}_{\Omega
}^{\dagger}$ a Rindler particle. The number of the Rindler particle in the
Minkowski vacuum is
\begin{align}
\left\langle \hat{N}_{\Omega}\right\rangle  &  =\left\langle 0_{M}\right\vert
\hat{b}_{\Omega}^{\dagger}\hat{b}_{\Omega}\left\vert 0_{M}\right\rangle
\nonumber\\
&  =\frac{1}{e^{\frac{2\pi\Omega}{a}}-1}\delta\left(  0\right)  ,
\end{align}
where $\delta\left(  0\right)  $ is the volume $V$ of the space
\cite{mukhanov2007introduction}. The density of the particle is then
\begin{equation}
\rho=\frac{\left\langle \hat{N}_{\Omega}\right\rangle }{V}=\frac{1}%
{e^{\frac{2\pi\Omega}{a}}-1}.
\end{equation}
This is a thermal distribution with the temperature
\begin{equation}
T=\frac{2\pi}{a}.
\end{equation}

For simplicity, we just consider the thermal radiation of the black hole with
a massless scalar field. The radiation of other species of particles can be
considered with the same procedure. For the black hole entropy, it is
necessary to consider the radiation of all species of particles
\cite{chen2018why}.

\section{Constructing white hole in Minkowski spacetime}

\subsection{White hole}

Rather than the thermal radiation, we show that for a white hole there is a
thermal absorption.

The white hole is defined as a region \cite{wald2010general}
\begin{equation}
W=M-J^{+}\left(  \mathcal{J}^{-}\right)  ,
\end{equation}
where $M$ is the whole spacetime, $\mathcal{J}^{-}$ is the past null infinity,
and $J^{+}\left(  \mathcal{J}^{-}\right)  $ is the causal future of
$\mathcal{J}^{-}$.

Similarly as that in constructing the black hole, the white hole is
constructed by setting a boundary%
\begin{equation}
t=-\sqrt{x^{2}+\eta^{2}},\text{ }t<0 \label{WH.Bound}%
\end{equation}
in the 1+1-dimensional Minkowski spacetime, eq. (\ref{Metric}). The boundary
(\ref{WH.Bound}) is the spacetime singularity.

\begin{figure}[h]
\centering
\includegraphics[width=9.0cm,height=6.0cm]{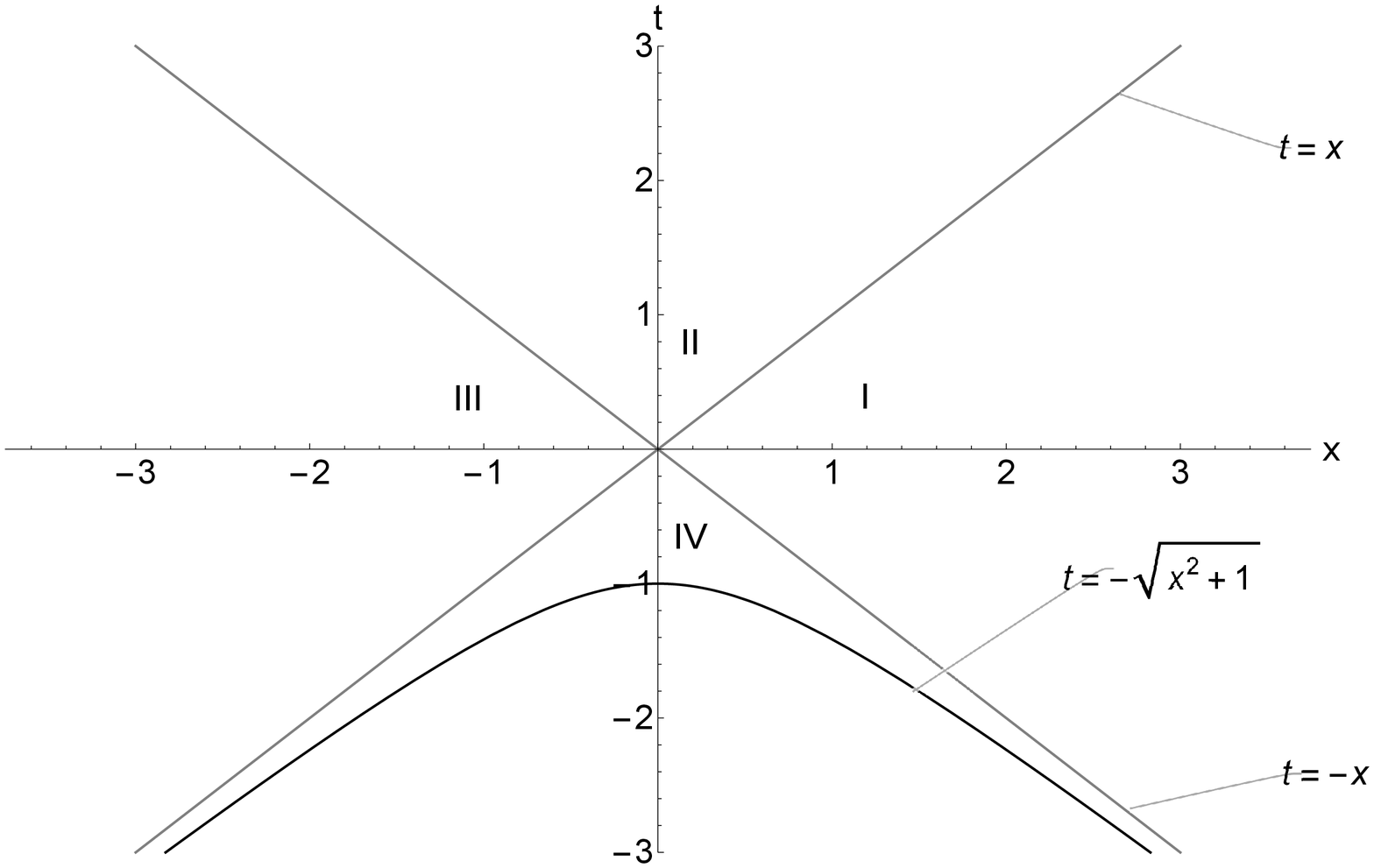}
\caption{The toy white hole}
\end{figure}

\subsection{Thermal absorption}

In order to consider the thermal absorption of the white hole, we consider a
massless scalar field $\phi$ in eq. (\ref{Massless.KG}) in the white hole
spacetime. The general solution is eq. (\ref{Solution.general}).

The region IV in fig. 3 is the region with $t-x<0$ and $t+x<0$.\textbf{ }In
the region IV we introduce new coordinates,
\begin{align}
t  &  =-r\cosh\theta,\nonumber\\
x  &  =r\sinh\theta, \label{WH.coordinate}%
\end{align}
and then the boundary (\ref{WH.Bound}) becomes%
\begin{equation}
r=\eta. \label{WH.Boundary}%
\end{equation}
We choose the boundary condition
\begin{equation}
\left.  \frac{\partial\phi}{\partial r}\right\vert _{r=\eta}=0 \label{WH.BC.2}%
\end{equation}
for the white hole boundary (\ref{WH.Bound}) or (\ref{WH.Boundary}).

Using the coordinate (\ref{WH.coordinate}) and substituting eq.
(\ref{Solution.general}) into the boundary condition (\ref{WH.BC.2}), we have%
\begin{equation}
f^{\prime}\left(  -\eta e^{\theta}\right)  =-e^{-2\theta}h^{\prime}\left(
-\eta e^{-\theta}\right)  . \label{assistant.2}%
\end{equation}
Integrating eq. (\ref{assistant.2}) over $-\eta e^{\theta}$,
\begin{align}
\int_{\theta=\theta_{0}}^{\theta}d\left(  -\eta e^{\theta}\right)  f^{\prime
}\left(  -\eta e^{\theta}\right)   &  =-\int_{\theta=\theta_{0}}^{\theta
}d\left(  -\eta e^{\theta}\right)  e^{-2\theta}h^{\prime}\left(  -\eta
e^{-\theta}\right)  ,\nonumber\\
&  =\int_{\theta=\theta_{0}}^{\theta}d\left(  -\eta e^{-\theta}\right)
f^{\prime}\left(  -\eta e^{-\theta}\right)  ,
\end{align}
we have
\begin{equation}
h\left(  -\eta e^{\theta}\right)  =f\left(  -\eta e^{-\theta}\right)  +C,
\end{equation}
with $C$ the integration constant. By eq. (\ref{WH.coordinate}) we have
\begin{align}
h\left(  -\eta\sqrt{\frac{t-x}{t+x}}\right)   &  =f\left(  -\eta\sqrt
{\frac{t+x}{t-x}}\right)  +C,\nonumber\\
h\left(  z\right)   &  =f\left(  \frac{\eta^{2}}{z}\right)  +C,
\end{align}
where $z\equiv-\eta\sqrt{\frac{t-x}{t+x}}<0$. Without loss of generality, we
choose $C=0$. This implies that the boundary condition (\ref{WH.BC.2})
requires
\begin{equation}
h\left(  z\right)  =f\left(  \frac{\eta^{2}}{z}\right)  ,\text{ \ }z<0.
\label{relation.WH.BC.2}%
\end{equation}

In the region IV $t-x<0$ and $t+x<0$, so $f\left(  t-x\right)  $ and $h\left(
t+x\right)  $ must satisfy the relation (\ref{relation.WH.BC.2}). This means
that each in-going mode $h\left(  t+x\right)  $ equals to the out-going mode
$f\left(  \frac{\eta^{2}}{t+x}\right)  $. When $f\left(  t-x\right)  =0$, we
have $h\left(  t+x\right)  =f\left(  \frac{\eta^{2}}{t+x}\right)  =0$.

In the region I, however, the situation changes. In the region I $t+x>0$ and
$t-x<0$, $f\left(  t-x\right)  $ is also restrained by the relation
(\ref{relation.WH.BC.2}) while $h\left(  t+x\right)  $ is no longer restrained
by the relation (\ref{relation.WH.BC.2}). That is, even though $f\left(
t-x\right)  =0$, $h\left(  t+x\right)  $ may still be nonvanishing. This means
that the number of the in-going modes is more than the number of the out-going
modes. The extra part of the in-going mode comes from the past null infinite
and goes to the white hole horizon, which is the absorption of the white hole.
By the same calculation in the previous section, the extra part of the
in-going mode satisfies a thermal distribution%
\begin{equation}
\left\langle \hat{N}_{\Omega}\right\rangle =\frac{V}{e^{\frac{2\pi\Omega}{a}%
}-1}.
\end{equation}
That is, the white hole horizon has a thermal absorption.

\section{Constructing wormhole in Minkowski spacetime}

In this section, we construct a wormhole which can also be regarded as a
spacetime in which there coexist a black hole and a white hole. For example,
the maximal manifold of the Schwarzschild spacetime contains an Einstein-Rosen
bridge which is also call a wormhole \cite{ohanian2013gravitation}.

The wormhole\ is constructed by setting the boundary%
\begin{equation}
t^{2}-x^{2}=\eta^{2} \label{WM.Bound}%
\end{equation}
in the Minkowski spacetime, fig. 4.

\begin{figure}[h]
\centering
\includegraphics[width=9.0cm,height=6.0cm]{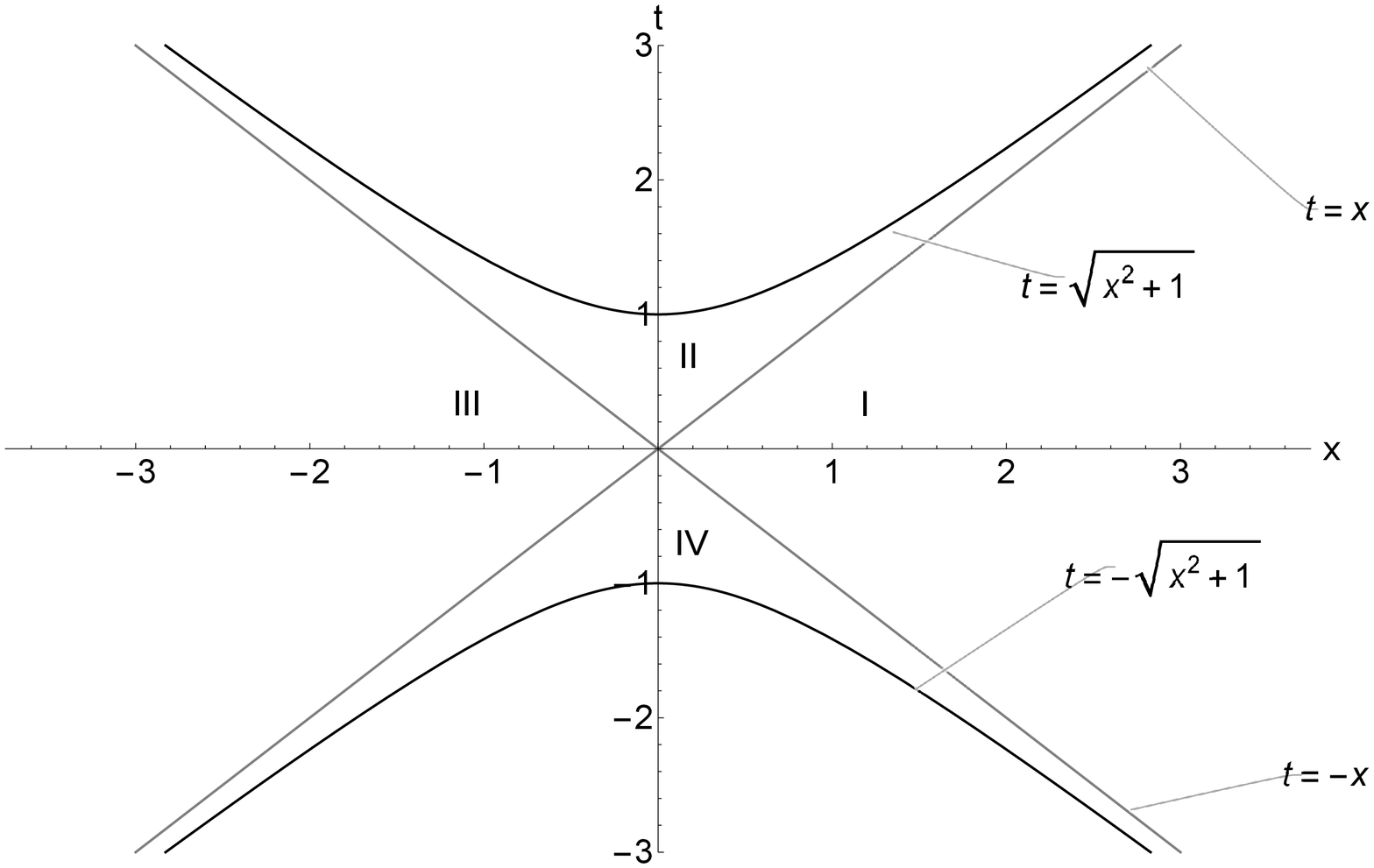}
\caption{A toy wormhole}
\end{figure}

With the boundary condition (\ref{BH.BC.2}) at $t=\sqrt{x^{2}+\eta^{2}}$ and
the boundary condition (\ref{WH.BC.2}) at $t=-\sqrt{x^{2}+\eta^{2}}$, we
obtain the solution of eq. (\ref{Massless.KG})
\begin{equation}
\phi=f\left(  t-x\right)  +f\left(  \frac{\eta^{2}}{t+x}\right)  .
\label{WM.solution}%
\end{equation}

In the whole spacetime, including the regions I, II, III, and IV, the in-going
modes are consistent with the out-going modes. That is, in the toy
wormhole\textbf{ }spacetime, the radiation and the absorption counteracts.

For black holes, there exists a information loss problem. In the case of black
holes, the extra out-going modes outside the horizon is explained as a
radiation from a black hole. Because the extra out-going modes satisfy the
thermal distribution and deliver no information; so, when a black hole
evaporates through the thermal radiation, the information disappears. This is
the so called information loss problem. When a black hole and a white whole
coexist, as analyzed above, there are no extra modes outside the horizon. That
is, there is no information loss for the spacetime in which a black hole and a
white hole coexist.

The maximum manifold of the Schwarzschild spacetime has the same causal
structure with the spacetime shown in fig. 3, in which a black hole and a
white hole coexist. If we suppose that the argument above is also available
for the physical picture in the maximum manifold of the Schwarzschild
spacetime, we come to a conclusion that in the maximum manifold of the
Schwarzschild spacetime, there is no information loss.

In addition, one may consider more complicated boundary conditions for the
wormhole\textbf{ }spacetime. For example, one may mathematically choose the
boundary condition (\ref{BH.BC.1}) in the later section for the black hole and
the boundary condition (\ref{WH.BC.2}) for the white hole. This will make the
problem more difficult. We choose the same boundary condition at the black
hole singularity and the white hole singularity for two reasons: (1) It makes
the problem easy to solve mathematically. (2) In the maximum manifold of the
Schwarzschild solution, the singularity in the black hole and the singularity
in the white hole are both $r=0$\textbf{, }so that it is physically natural to
choose the same boundary condition for them.

On the other hand, we notice that when $t=0$,
\begin{equation}
\phi\left(  x\right)  =\phi\left(  -\frac{\eta^{2}}{x}\right)  ,
\end{equation}
which means that the region $x>0$ and $x<0$ are dual to each other. Once an
observer detects the field in one side, the observer knows the field in the
other side. If we believe that all the information of the system are contained
in the field, then the region $x>0$ and the region $x<0$ possess the same
information. No matter the wormhole is transversible or not, an observer in
one side can know all the information in the other side. We suppose this
property is independent of the symmetry of the spacetime.

\section{Effect caused by boundary condition}

In this section, we discuss the physical picture of two different boundary conditions.

Taking the black hole in fig. 1 as an example. In the case of black holes, the
radiation is independent of the choice of boundary conditions. To illustrate,
we choose another boundary condition
\begin{equation}
\left.  \phi\right\vert _{r=\eta}=0. \label{BH.BC.1}%
\end{equation}
In the boundary condition (\ref{BH.BC.1}), $r$ is defined in eq.
(\ref{BH.coordinate}) and $r=\eta$ is the branch of the hyperbolic curve
$t^{2}-x^{2}=\eta^{2}$ above the $x$-axis. With the boundary condition
(\ref{BH.BC.1}), we have
\begin{equation}
f\left(  z\right)  =-h\left(  \frac{\eta^{2}}{z}\right)  ,\text{ \ }z>0.
\label{relation.BH.BC.1}%
\end{equation}
With the same analysis of the boundary condition (\ref{BH.BC.2}), the black
hole also radiates thermally. That is, the thermal radiation of the black hole
is independent of the choice of the boundary condition.

By the same procedure, in the case of the white hole, we choose another
boundary condition
\begin{equation}
\left.  \phi\right\vert _{r=\eta}=0. \label{WH.BC.1}%
\end{equation}
In the boundary condition (\ref{WH.BC.1}), $r$ is defined in eq.
(\ref{WH.coordinate}) and $r=\eta$ is the branch of the hyperbolic curve
$t^{2}-x^{2}=\eta^{2}$ below the $x$-axis. By the same analysis of the
boundary condition (\ref{WH.BC.2}), the white hole also absorbs thermally.
That is, the thermal absorption of the white whole is independent of the
choice of boundary conditions.

Mathematically, the boundary conditions (\ref{BH.BC.1}) and (\ref{WH.BC.1})
are the Dirichlet boundary conditions and the boundary conditions
(\ref{BH.BC.2}) and (\ref{WH.BC.2}) are the von Newmann boundary conditions.
From the view of physics, the boundary condition (\ref{BH.BC.1}) implies the
reflection on the boundary and the boundary conditions (\ref{BH.BC.2}) implies
the absorption on the boundary.

With the boundary condition (\ref{WH.BC.2}), the solution in the region II is
\begin{equation}
\phi=f\left(  t-x\right)  +f\left(  \frac{\eta^{2}}{t+x}\right)  .
\label{BH.Sol.2}%
\end{equation}
Take the mode expansion of $\phi$:
\begin{equation}
\phi=\int_{-\infty}^{\infty}dk\left(  a_{k}e^{ik\left(  t-x\right)  }%
+a_{k}^{\dagger}e^{-ik\left(  t-x\right)  }\right)  +\int_{-\infty}^{\infty
}dk\left(  b_{k}e^{ik\left(  t+x\right)  }+b_{k}^{\dagger}e^{-ik\left(
t+x\right)  }\right)  . \label{BH.Mod.2}%
\end{equation}
Here, $a_{k}$ and $a_{k}^{\dagger}$ are the creation operator and annihilation
operator of the out-going modes; $b_{k}$ and $b_{k}^{\dagger}$ are the
creation operator and annihilation operator of the in-going modes. Generally
\[
f\left(  t+x\right)  \neq f\left(  \frac{\eta^{2}}{t+x}\right)  ,
\]
so that
\[
a_{k}^{\dagger}\neq b_{k}^{\dagger}.
\]
That is, an in-going particle created by $b_{k}^{\dagger}$ and an out-going
particle created by $a_{k}^{\dagger}$ are two different kinds of particles.

With the boundary condition (\ref{WH.BC.1}), the solution in region II is
\begin{equation}
\phi=f\left(  t-x\right)  -f\left(  \frac{\eta^{2}}{t+x}\right)  .
\label{BH.Sol.1}%
\end{equation}
The mode expansion is
\begin{equation}
\phi=\int_{-\infty}^{\infty}dk\left(  a_{k}e^{ik\left(  t-x\right)  }%
+a_{k}^{\dagger}e^{-ik\left(  t-x\right)  }\right)  +\int_{-\infty}^{\infty
}dk\left(  b_{k}^{\prime}e^{ik\left(  t+x\right)  }+b_{k}^{\prime\dagger
}e^{-ik\left(  t+x\right)  }\right)  . \label{BH.Mod.1}%
\end{equation}
Here $a_{k}$ and $a_{k}^{\dagger}$ are the creation and annihilation operators
of the out-going particle, $b_{k}^{\prime}$ and $b_{k}^{\prime\dagger}$ are
the creation and annihilation operators of the out-going particle. Generally
\[
f\left(  t+x\right)  \neq-f\left(  \frac{\eta^{2}}{t+x}\right)  ,
\]
so that
\[
a_{k}^{\dagger}\neq b_{k}^{\prime\dagger}.
\]
That is, an in-going particle created by $b_{k}^{\prime\dagger}$ and an
out-going particle created by $a_{k}^{\dagger}$ are two different kinds of
particles.\textbf{ }Comparing eqs. (\ref{BH.Sol.2}) and (\ref{BH.Sol.1}) we
have%
\[
b_{k}^{\prime\dagger}=b_{k}^{\dagger}e^{i\pi}.
\]
The in-going mode with the reflection boundary condition (\ref{BH.BC.1}) has a
half-wave loss comparing with the in-going mode with the absorption boundary
condition (\ref{BH.BC.2}).

\section{Conclusion and outlook}

The black hole and the white hole are universal structures of the spacetime.
It is irrelevant with the details of the local metric structure of the
spacetime. When considering a quantum field on a black hole spacetime or a
white hole spacetime, we should concern the universal property rather than the
local property.

In this paper, we build toy models of the black hole, the white hole, and the
wormhole (a spacetime in which a black hole and a white hole coexist). The
models are achieved by setting proper boundaries in the Minkowski spacetime.
The advantage of this toy models is that the local property of the spacetime
background is simple enough to deal with. This allows us to concentrate on the
universe property of the quantum field in these models.

We show the thermal radiation of the toy black hole. By counting the number of
the modes, we find that outside the black hole there exist more out-going
modes than in-going modes and the extra out-going modes satisfies a thermal
distribution. These extra out-going modes outside the black hole or the
horizon are the thermal radiation of the black hole. Similarly, we show that
the white hole has a thermal absorption. That is, outside the white hole,
there exist more in-going modes than out-going modes and the extra in-going
modes satisfy a thermal distribution. We show that there is no information
loss in the wormhole spacetime. We exemplify that the black hole radiation and
the white hole absorption are independent of the choice of the boundary
condition, by comparing the radiation and the absorption with the Dirichlet
boundary condition and the von Newmann boundary condition. The Dirichlet
boundary causes half-wave loss of the in-going mode comparing with the von
Newmann boundary condition.

%\appendix
%\section{Some title}
%Please always give a title also for appendices.

\acknowledgments

We are very indebted to Dr G. Zeitrauman for his encouragement. This work is
supported in part by Nankai Zhide foundation and NSF of China under Grant No.
11575125 and No.11675119.

%(正文结束)――――――――――――――――――――――――――――――――――――――――――――――――――

%%%%%%%%%%%%%%%%%%%参考文献%%%%%%%%%%%%%%%%%%%%%%%%%%%

%\begin{thebibliography}{10}

%bbl放在这
\providecommand{\href}[2]{#2}\begingroup\raggedright\endgroup

%\end{thebibliography}

%%%%%%%%%%%%%%%%%%%bibtex形式的参考文献%%%%%%%%%%%%%%%
%\bibliographystyle{JHEP} %参考文献的风格(.bst)
%\bibliography{refs} %参考文献文件(.bib)

\begin{thebibliography}{10}

\bibitem{hawking1975particle}
S.~W. Hawking, {\it Particle creation by black holes},  {\em Communications in
  mathematical physics} {\bf 43} (1975), no.~3 199--220.

\bibitem{boulware1976hawking}
D.~G. Boulware, {\it Hawking radiation and thin shells},  {\em Physical Review
  D} {\bf 13} (1976), no.~8 2169.

\bibitem{davies1976origin}
P.~Davies, {\it On the origin of black hole evaporation radiation},  {\em
  Proceedings of the Royal Society of London. A. Mathematical and Physical
  Sciences} {\bf 351} (1976), no.~1664 129--139.

\bibitem{zhang2005new}
J.~Zhang and Z.~Zhao, {\it New coordinates for kerr--newman black hole
  radiation},  {\em Physics Letters B} {\bf 618} (2005), no.~1-4 14--22.

\bibitem{majhi2010hawking}
B.~R. Majhi, {\it Hawking radiation and black hole spectroscopy in
  hovrava--lifshitz gravity},  {\em Physics Letters B} {\bf 686} (2010), no.~1
  49--54.

\bibitem{konoplya2019quasinormal}
R.~Konoplya, A.~Zinhailo, and Z.~Stuchlik, {\it Quasinormal modes, scattering,
  and hawking radiation in the vicinity of an einstein-dilaton-gauss-bonnet
  black hole},  {\em Physical Review D} {\bf 99} (2019), no.~12 124042.

\bibitem{li2008hawking}
R.~Li, J.-R. Ren, and S.-W. Wei, {\it Hawking radiation of dirac particles via
  tunneling from the kerr black hole},  {\em Classical and Quantum Gravity}
  {\bf 25} (2008), no.~12 125016.

\bibitem{sakalli2015hawking}
I.~Sakalli and A.~Ovgun, {\it Hawking radiation of spin-1 particles from a
  three-dimensional rotating hairy black hole},  {\em Journal of Experimental
  and Theoretical Physics} {\bf 121} (2015), no.~3 404--407.

\bibitem{sakalli2016black}
I.~Sakalli and A.~Ovgun, {\it Black hole radiation of massive spin-2 particles
  in (3+1) dimensions},  {\em The European Physical Journal Plus} {\bf 131}
  (2016), no.~6 1--13.

\bibitem{volovik2016black}
G.~E. Volovik, {\it Black hole and hawking radiation by type-ii weyl fermions},
   {\em JETP letters} {\bf 104} (2016), no.~9 645--648.

\bibitem{parker1975probability}
L.~Parker, {\it Probability distribution of particles created by a black hole},
   {\em Physical Review D} {\bf 12} (1975), no.~6 1519.

\bibitem{wald1975particle}
R.~M. Wald, {\it On particle creation by black holes},  {\em Communications in
  Mathematical Physics} {\bf 45} (1975), no.~1 9--34.

\bibitem{hawking1976black}
S.~W. Hawking, {\it Black holes and thermodynamics},  {\em Physical Review D}
  {\bf 13} (1976), no.~2 191.

\bibitem{hartle1976path}
J.~B. Hartle and S.~W. Hawking, {\it Path-integral derivation of black-hole
  radiance},  {\em Physical Review D} {\bf 13} (1976), no.~8 2188.

\bibitem{damour1976black}
T.~Damour and R.~Ruffini, {\it Black-hole evaporation in the
  klein-sauter-heisenberg-euler formalism},  {\em Physical Review D} {\bf 14}
  (1976), no.~2 332.

\bibitem{banerjee2009hawking}
R.~Banerjee and B.~R. Majhi, {\it Hawking black body spectrum from tunneling
  mechanism},  {\em Physics Letters B} {\bf 675} (2009), no.~2 243--245.

\bibitem{steinhauer2014observation}
J.~Steinhauer, {\it Observation of self-amplifying hawking radiation in an
  analogue black-hole laser},  {\em Nature Physics} {\bf 10} (2014), no.~11
  864--869.

\bibitem{steinhauer2016observation}
J.~Steinhauer, {\it Observation of quantum hawking radiation and its
  entanglement in an analogue black hole},  {\em Nature Physics} {\bf 12}
  (2016), no.~10 959--965.

\bibitem{de2019observation}
J.~R.~M. de~Nova, K.~Golubkov, V.~I. Kolobov, and J.~Steinhauer, {\it
  Observation of thermal hawking radiation and its temperature in an analogue
  black hole},  {\em Nature} {\bf 569} (2019), no.~7758 688--691.

\bibitem{macher2009black}
J.~Macher and R.~Parentani, {\it Black-hole radiation in bose-einstein
  condensates},  {\em Physical Review A} {\bf 80} (2009), no.~4 043601.

\bibitem{hawking2015information}
S.~W. Hawking, {\it The information paradox for black holes},  {\em
  arXiv:1509.01147} (2015).

\bibitem{polchinski2015new}
J.~Polchinski, {\it The black hole information problem},  {\em
  arXiv:1609.04036v1} (2016).

\bibitem{li2019scattering}
W.-D. Li, Y.-Z. Chen, and W.-S. Dai, {\it Scattering state and bound state of
  scalar field in schwarzschild spacetime: Exact solution},  {\em Annals of
  Physics} {\bf 409} (2019) 167919.

\bibitem{li2018scalar}
W.-D. Li, Y.-Z. Chen, and W.-S. Dai, {\it Scalar scattering in schwarzschild
  spacetime: Integral equation method},  {\em Physics Letters B} {\bf 786}
  (2018) 300--304.

\bibitem{wald2010general}
R.~M. Wald, {\em General relativity}.
\newblock University of Chicago press, 2010.

\bibitem{ohanian2013gravitation}
H.~C. Ohanian and R.~Ruffini, {\em Gravitation and spacetime}.
\newblock Cambridge University Press, 2013.

\bibitem{mukhanov2007introduction}
V.~Mukhanov and S.~Winitzki, {\em Introduction to quantum effects in gravity}.
\newblock Cambridge university press, 2007.

\bibitem{susskind2005introduction}
L.~Susskind and J.~Lindesay, {\em An introduction to black holes, information
  and the string theory revolution: The holographic universe}.
\newblock World Scientific, 2005.

\bibitem{chen2018why}
Y.~Chen, W.~Li, and W.~Dai, {\it Why the entropy of spacetime is independent of
  species of particles: the species problem},  {\em European Physical Journal
  C} {\bf 78} (2018), no.~8 635.

\end{thebibliography}
%\nocite{*} %若去掉注释，没有被引用的文献也被列出

%%%%%%%%%%%%%%%%%%%bbl形式的参考文献%%%%%%%%%%%%%%%%%%

%%%%%%%%%%%%%%%%%%%%%%%%%%%%%%%%%%%%%%%%%%%%%%%%%%%%%%

\end{CJK*}
\end{document}